\documentclass{iopart}
\usepackage{epsfig}

\newcommand{\hL}{\hat{\mathbf{L}}}
\newcommand{\hS}{\hat{\mathbf{S}}}

\newcommand{\vS}{\mathbf{S}}
\newcommand{\vL}{\mathbf{L}}
\newcommand{\vn}{\mathbf{n}}

\newcommand{\beq}{\begin{equation}}
\newcommand{\eeq}{\end{equation}}
\newcommand{\bes}{\begin{subequations}}
\newcommand{\ees}{\end{subequations}}
\newcommand{\bea}{\begin{eqnarray}}
\newcommand{\eea}{\end{eqnarray}}
\newcommand{\ba}{\begin{array}}
\newcommand{\ea}{\end{array}}
\newcommand{\beqn}{\begin{eqnarray*}}
\newcommand{\eeqn}{\end{eqnarray*}}

\newcommand{\lisa}{{\em LISA}}

\def\bt{\bar \theta}
\def\bph{\bar \phi}

\def\nn{\nonumber}

\def\ii{{\rm i}}   

\begin{document}

\title[Testing general relativity and probing massive black hole
mergers with LISA]{Testing general relativity and probing the merger
history of massive black holes with LISA}

\author{Emanuele Berti\footnote[3]{To whom correspondence should be
addressed (berti@wugrav.wustl.edu)},\dag\ Alessandra Buonanno\footnote[2]
{UMR 7164 (CNRS, Universit\'e Paris7, CEA, Observatoire de Paris).}
\ddag\ and Clifford M. Will\dag}

\address{\dag\ McDonnell Center for the Space Sciences, Department of
Physics, Washington University, St. Louis, Missouri 63130}

\address{\ddag\ AstroParticule et Cosmologie (APC),
11 place Marcelin Berthelot, 75231 Paris 
and Institut d'Astrophysique de Paris, 98$^{\rm bis}$ Boulv. Arago, 
75013 Paris, France}

\begin{abstract}
Observations of binary inspirals with the proposed Laser
Interferometer Space Antenna ({\em LISA}) will allow us to place bounds
on alternative theories of gravity and to study the merger history of
massive black holes (MBH). These possibilities rely on \lisa's
parameter estimation accuracy. We update previous studies of
parameter estimation for inspiralling compact binaries of MBHs, and
for inspirals of neutron stars into intermediate-mass black holes,
including non-precessional spin effects. We work both in
Einstein's theory and in alternative theories of gravity of the
scalar-tensor and massive-graviton types. Inclusion of 
non-precessional spin terms in MBH binaries has little effect on
the angular resolution or on distance determination accuracy, but it
degrades the estimation of intrinsic binary parameters such as chirp
mass and reduced mass by between one and two orders of magnitude.  The
bound on the coupling parameter $\omega_{\rm BD}$ of scalar-tensor
gravity is significantly reduced by the presence of spin couplings,
while the reduction in the graviton-mass bound is milder.  {\em LISA}
will measure the luminosity distance of MBHs to better than $\sim
10\%$ out to $z\simeq 4$ for a $(10^6+10^6)~M_\odot$ binary, and out
to $z\simeq 2$ for a $(10^7+10^7)~M_\odot$ binary. The chirp mass of a
MBH binary can always be determined with excellent accuracy. Ignoring
spin effects, the reduced mass can be measured within $\sim 1~\%$ out
to $z=10$ and beyond for a $(10^6+10^6)~M_\odot$ binary, but only out
to $z\simeq 2$ for a $(10^7+10^7)~M_\odot$ binary. Present-day MBH
coalescence rate calculations indicate that most detectable events
should originate at $z\sim 2-6$: at these redshifts {\em LISA} can be
used to measure the two black hole masses and their luminosity
distance with sufficient accuracy to probe the merger history of
MBHs. If the low-frequency {\em LISA} noise can only be trusted down
to $10^{-4}$~Hz, parameter estimation for MBHs (and {\em LISA}'s
ability to perform reliable cosmological observations) will be
significantly degraded.

\end{abstract}

\maketitle

\section{Summary of results}

The Laser Interferometer Space Antenna ({\em LISA}) is being designed
to detect gravitational wave (GW) signals in the frequency band
between $10^{-4}$~Hz and $10^{-1}$~Hz~\cite{danzmann}. Operating at
these low frequencies, {\em LISA} can detect, among other sources, the
inspiral of stellar-mass compact objects -- such as neutron stars (NS)
or black holes (BH) -- into intermediate-mass black holes (IMBH) with
masses in the range $10^2 \mbox{--} 10^4 M_\odot$.  Another strong
gravitational wave source to be observed by {\em LISA} is the inspiral
and merger of massive black holes (MBH) with masses in the range $10^4
\mbox{--} 10^7 M_\odot$.

Gravitational radiation reaction drives the inspiral of stellar-mass
compact objects into IMBHs. It also dominates the final stages in the
evolution of coalescing MBH binaries.  The amplitude and phase of the
gravitational wave signal carry information about binary parameters,
such as masses and spins, and about the location and distance of the
binary.  They may also be different in different theories of gravity.
Therefore {\em LISA} can provide important astrophysical information,
yield interesting tests of fundamental physics, and place bounds on
alternative theories of gravity.  In this paper,
we consider, along with
standard general relativity, theories of the scalar-tensor type (the
simplest exemplar being that of Brans and Dicke) and theories with an
effective mass in the propagation of gravitational waves (which we
call massive graviton theories, for short).  In scalar-tensor theories
the phasing evolution is modified predominantly by the presence of
dipole gravitational radiation reaction in the orbital evolution (in
general relativity the lowest radiative multipole moment is the
quadrupole).  In massive graviton theories the gravitational wave
propagation speed depends on wavelength: this generates a distortion
in the time of arrival (and in the wave phasing) with respect to
general relativity, similar to the dispersion of radio waves by
interstellar plasma.

In a recent paper \cite{BBW} (henceforth BBW) we investigated the
effect of spin-orbit and spin-spin couplings both on the estimation of
astrophysical parameters within general relativity, and on bounds that
can be placed on alternative theories.  We restricted our analysis to
non-precessing spinning binaries, i.e.  binaries whose spins are 
parallel (or anti-parallel)
to the orbital angular momentum. Our work extended
results of previous papers~\cite{willST,willgraviton,scharrewill,willyunes,damourfarese}
that derived bounds on the graviton mass, on the Brans-Dicke parameter
$\omega_{\rm BD}$ and on parameters describing more general
scalar-tensor theories under the assumption that the compact objects
do not carry spin.

Within Einstein's general relativity, various authors have
investigated the accuracy with which {\em LISA} can determine binary
parameters including spin effects~\cite{CC,SH,seto,vecchio,HM}. 
These earlier works (except for~\cite{HM}) adopted analytical
approximations to {\em LISA}'s instrumental noise~\cite{CC}, augmented
by an estimate of white-dwarf confusion noise~\cite{BH} in the
low-frequency band.  In BBW (and in the present work) we model the
{\em LISA} noise curve by a similar -- albeit slightly updated --
analytical approximation (\cite{BC}; see Sec.~IIC and Fig.~1 of BBW
for details).  This noise curve has the advantage of being given in
analytical form, and reproduces very well the salient features of
numerical noise curves available online from the {\em LISA}
Sensitivity Curve Generator~\cite{SCG}, a tool sponsored by the {\em
LISA} International Science Team.

The central conclusions of our work are as follows.  Inclusion of
non-precessing spin-orbit and spin-spin terms in the gravitational
wave phasing generally reduces the accuracy with which the parameters
of the binary can be estimated.  This is not surprising, since the
parameters are highly correlated, and adding parameters effectively
dilutes the available information.  Such an effect has already been
described within Einstein's general relativity in the context of
ground-based detectors of the LIGO/VIRGO
type~\cite{poissonwill,KKS}. For example, for massive black-hole
binaries at 3 Gpc, we find that including spin-orbit terms degrades
the accuracy in measuring chirp mass by factors of order 10, and in
measuring the reduced mass parameter by factors of order 20 -- 100;
including spin-spin terms further degrades these accuracies by factors
of order 3 and 5, respectively.  For neutron stars inpiralling into
IMBHs with masses between $10^3$ and $10^4$ solar masses, the
corresponding reductions are factors of order 20 and 5 -- 30 in chirp
mass and reduced mass parameter, respectively, when spin-orbit is
included, and additional factors of order 4 and 7, respectively, when
spin-spin terms are included.

When we consider placing bounds on alternative theories of gravity,
for technical reasons, we treat only spin-orbit terms.  The source of
choice to place bounds on the coupling parameter $\omega_{\rm BD}$ of
scalar-tensor gravity is the inspiral of a neutron star into an
IMBH. We first reproduce results of earlier work~\cite{willyunes},
apart from small differences arising from corrected normalization of
the \lisa ~noise curve.  Then, we include spin-orbit effects and find 
that the bound on $\omega_{\rm BD}$ is significantly reduced, 
by factors of order 10 --20.  
For example for a $1.4 \, M_\odot$ neutron star inspiralling into
a $400 \, M_\odot$ BH, the average bound on $\omega_{\rm BD}$ from a
population of binaries across the sky goes from $3\times 10^5$ to
$2 \times 10^4$ when spin-orbit terms are included.  The latter (average)
bound should be compared with the bound of $4 \times 10^4$ from {\em Cassini}
measurements of the Shapiro time delay~\cite{bertotti}.

The effect of including spin on bounding the graviton mass is more
modest. In this case, the source of choice is the inspiral of MBH
binaries.  For masses ranging from $10^5 M_\odot$ to $10^7 M_\odot$,
the reduction in the bound induced by the inclusion of spin-orbit
terms is only a factor of 4 to 5.

We also consider the effect of spin terms on the angular and distance
resolution of \lisa. We find that spin couplings have a mild effect on
the angular resolution, on the distance and, as a consequence, on the
redshift determination for MBH binaries. By contrast, for stellar mass
objects inspiralling into IMBHs, neither distance nor location on the
sky is very well determined. Furthermore, for NS-IMBH binaries the
angular resolution is somewhat dependent on the inclusion of spin
terms. Here we show that the different spin dependence of the angular
resolution in the NS-IMBH and MBH-MBH cases can be traced back to a
different behavior of the correlations between the chirp mass (leading
the gravitational wave evolution of the binary) and the angular
variables describing {\lisa}'s position in the sky.

{\lisa}'s low-frequency sensitivity affects the accuracy of estimating
parameters for MBH binaries, as we showed in BBW (similar investigations
can be found in~\cite{HH} and~\cite{Baker}). As default low-frequency
cutoff we choose $10^{-5}$~Hz. However, below $10^{-4}$~Hz, the noise
characteristics of \lisa~are uncertain, and if we set {\lisa}'s
low-frequency cutoff at $10^{-4}$~Hz, the accuracy of estimating both
extrinsic parameters such as distance and sky position and intrinsic
parameters such as chirp mass and reduced mass, as well as the
graviton mass, can be significantly lower, especially for higher-mass
systems.

{\em LISA} can observe MBH binaries with large SNR out to large values
of the cosmological redshift. If the corresponding mass and distance
determinations are accurate enough, {\em LISA} will be a useful tool
to study BH formation in the early Universe. Using Monte Carlo
simulations we find that {\em LISA} can provide accurate distance
determinations out to redshift $z\sim 2$ for source masses $\sim 10^7
M_\odot$, and out to $z\sim 4$ for source masses $\sim 10^6
M_\odot$. Mass determinations strongly depend on an accurate treatment
of spin effects. The chirp mass of a MBH binary can always be
determined with excellent accuracy. Ignoring spin effects, the reduced
mass can be measured within $\sim 1~\%$ out to $z=10$ and beyond for a
$(10^6+10^6)~M_\odot$ binary, but only out to $z\simeq 2$ for a
$(10^7+10^7)~M_\odot$ binary. Present-day MBH coalescence rate
calculations indicate that most detectable events should originate at
$z\sim 2-6$ \cite{sesana2,RW,menou}: at these redshifts {\em LISA} can
be used to measure the two BH masses and their luminosity distance
with sufficient accuracy to probe the merger history of MBHs.

The paper is organized as follows. In Sec.~\ref{BBWsummary} we briefly
summarize our procedure to study parameter estimation with \lisa. All
calculations presented here adopt the ``more realistic'' of the two
models described in BBW: we take into account the orbital motion of
the detector and use Monte Carlo simulations of a population of
sources across the sky. In Sec.~\ref{results} and~\ref{cosmology} we
extend and clarify some results presented in BBW. Sec.~\ref{results}
shows the dependence of parameter estimation and tests of alternative
theories of gravity on the binary mass (BBW only considered a few
representative values of the masses). Sec.~\ref{cosmology} discusses
the cosmological reach of \lisa~and its ability to probe the merger
history of MBHs, extending and complementing the discussion in
Sec.~IIIC of BBW.

\section{Gravitational wave detection and parameter estimation}
\label{BBWsummary}

We assume, as in \cite{CC}, that two independent Michelson outputs can
be constructed from the readouts of the three \lisa~arms if the noise
in the arms is totally symmetric (but see \cite{prince} for discussion of
the most general combinations of \lisa~ outputs, and their
sensitivities). Since the modulation of the measured signal due to
\lisa's motion occurs on timescales $\sim 1$~year (much larger than
the binary's orbital period), the Fourier transform of the signal can
be evaluated in the stationary phase approximation, with the result
\bea
\label{hSPA}
\tilde h_\alpha(f) = \frac{\sqrt{3}}{2}
\left[\frac{{\cal M}^{5/6}}{\sqrt{30}\pi^{2/3} D_{\rm L}}\right]
\left \{ \frac{5}{4} \frac{\tilde{A}_\alpha(t)}{f^{7/6}} \right \}
e^{\ii\bigl(\psi(f)-\varphi_{p,\alpha}(t)-\varphi_D(t)\bigr)}
\,.
\eea
Here $f$ is the frequency of the gravitational waves; ${\cal M} =
\eta^{3/5} M$ is the ``chirp'' mass, with $M=m_1+m_2$ and $\eta =
m_1m_2/M^2$; $D_{\rm L}$ is the luminosity distance to the source;
$\tilde{A}_{\alpha}(t) = \{[1+ (\hat \vL \cdot
\vn)^2]^2\,F_{\alpha}^{+\,2} + 4 (\hat \vL \cdot
\vn)^2\,F_{\alpha}^{\times\,2}\}^{1/2}$, where $\hat \vL$ is the
orbital angular momentum unit vector, $\vn$ is a unit vector in the
direction of the source on the sky, and the quantities
$F_{\alpha}^{+,\times}$ are the pattern functions for the two
equivalent Michelson detectors. The time variable $t=t(f)$ is given by
an expansion to 2PN order, including also the Brans-Dicke parameter
and the graviton-mass term (see BBW and \cite{CC} for details);
$\varphi_{p,\alpha}(t)$ is the waveform polarization phase and
$\varphi_{D}(t)$ the Doppler phase.

We have adopted the standard ``restricted post-Newtonian
approximation'' for the waveform, in which the amplitude is expressed
to the leading order in a post-Newtonian expansion (an expansion for
slow-motion, weak-field systems in powers of $v \sim (M/r)^{1/2} \sim
(\pi{\cal M}f)^{1/3}$), while the phasing $\psi(f)$, to which laser
interferometers are most sensitive, is expressed to the highest
post-Newtonian (PN) order reasonable for the problem at hand. For
binaries with spins aligned (or anti-aligned) and normal to the
orbital plane, this is a valid approximation because the amplitude
varies slowly (on a radiation reaction timescale) compared to the
orbital period.  But when the spins are not aligned, modulations of
the amplitude on a precession timescale must be included. 
Preliminary studies show that precession effectively decorrelates
parameters, improving parameter estimation in a significant
way~\cite{vecchio}. The study of precessional modulations is beyond
the scope of this work. 

The phasing function $\psi(f)$ is known for point masses up to 3.5 PN
order~\cite{blanchet1,blanchet2}.  But spin terms are known only up to
2PN order, so to be reasonably consistent, we include in the phasing
point-mass terms only up to this same 2PN order. The needed expression
for the phasing is
%
%
\bea
\psi(f)&=&2\pi f t_c-\phi_c
\\
&+&\frac{3}{128}\,(\pi {\cal M}f)^{-5/3}
\biggl\{1-
\frac{5 (\hat \alpha_1-\hat \alpha_2)^2}{336 \omega_{\rm BD}}\,\eta^{2/5}\,
(\pi {\cal M} f)^{-2/3}\,\nn
\\
&-&\frac{128}{3}\frac{\pi^2 D\,{\cal M}}{\lambda_g^2\,(1+z)}\,(\pi {\cal M} f)^{2/3}+\left(\frac{3715}{756}+\frac{55}{9}\eta\right)\,\eta^{-2/5}\,(\pi {\cal M} f)^{2/3}\nn\\
&-&\left.16\pi\,\eta^{-3/5}\,(\pi {\cal M} f) + 4 \beta\,\eta^{-3/5}\,(\pi {\cal M} f)
-10 \sigma\,\eta^{-4/5}\,(\pi {\cal M} f)^{4/3} \right.\nn\\
&+&\left. \left(\frac{15293365}{508032}+\frac{27145}{504}\eta+\frac{3085}{72}\eta^2\right)\,
\eta^{-4/5}\,(\pi {\cal M} f)^{4/3} 
\right\}\,,\nn
\eea
%
where $t_c$ and $\phi_c$ are the time and phase at coalescence. The
second term inside the braces is the contribution of dipole
gravitational radiation in Brans-Dicke theory: it depends on the
difference of the scalar charges of the two bodies $\alpha_i^2\simeq
\hat \alpha_i^2/2\omega_{BD}$ ($i=1,2$)~\footnote{Note that the
quantity $\hat{\alpha}_i = 1 - 2 s_i$, where $s_i$ is the so-called
sensitivity of the body~\cite{willST}. In calculations involving NSs
we adopt the value $\hat\alpha_{\rm NS} = 0.6$, a typical value
obtained from calculations of NS structure with realistic equations of
state in scalar-tensor theories \cite{willzaglauer}.}.

For black holes, $\alpha_i \equiv 0$, therefore we need ``mixed''
binaries (eg. NS-IMBH binaries) to place bounds on scalar-tensor
theories. The third term in the braces is the effect of a massive
graviton, which alters the arrival time of waves of a given frequency,
depending on the size of the graviton Compton wavelength $\lambda_g$
and on a distance quantity $D$ (see BBW or \cite{willgraviton} for the
relation between $D$ and the luminosity distance $D_L$).  
The remaining terms in the
braces are the standard general relativistic, post-Newtonian terms,
including spin effects.

The quantities $\beta$ and $\sigma$ represent spin-orbit and spin-spin
contributions to the phasing, given by
\bea 
\beta &=& \frac{1}{12} \sum_{i=1}^2 \chi_i \left [113 \frac{m_i^2}{M^2} + 75
\eta \right ] \hL \cdot \hS_i \,,
\\
\sigma &=& \frac{\eta}{48} \chi_1 \chi_2 \left (-247 \hS_1 \cdot \hS_2 + 721 \hL
\cdot \hS_1 \hL \cdot \hS_2 \right )\,,
\eea
where $\hS_i$ and $\hL$ are unit vectors in the direction of the spins
and of the orbital angular momentum, respectively, and $\vS_i=\chi_i
m_i^2 \hS_i$.  For BHs, the dimensionless spin parameters $\chi_i$
must be smaller than unity, while for neutron stars, they are
generally much smaller than unity. It follows that $|\beta|< 9.4$ and
$|\sigma|< 2.5$.

We use the by now standard machinery of parameter estimation in
matched filtering for gravitational wave detection that has been
developed by a number of
authors~\cite{poissonwill,finn,FinnChernoff,CutlerFlanagan}.  Given a
set of parameters $\theta^a$ characterizing the waveform, one defines
the ``Fisher matrix'' $\Gamma_{ab}$ with components given by
\begin{equation}
\Gamma_{ab} \equiv \left( \frac{\partial h}{\partial\theta^a} \mid
\frac{\partial
h}{\partial\theta^b} \right)
\equiv 2 \int_0^{\infty} 
\frac{1}{S_n(f)}
\left[{\frac{\partial \tilde h^*}{\partial\theta^a}}
{\frac{\partial \tilde h}{\partial\theta^b}}
+{{\frac{\partial \tilde h^*}{\partial\theta^b}}}
{\frac{\partial \tilde h}{\partial\theta^a}}\right]df\,.
\label{fisher}
\end{equation}
An estimate of the rms error, $\Delta\theta^a$, in measuring the
parameter $\theta^a$ can then be calculated, in the limit of large
SNR, by taking the square root of the diagonal elements of the inverse
of the Fisher matrix,
\begin{equation}
\Delta\theta^a = \sqrt{\Sigma^{aa}} \,, \qquad  \Sigma = \Gamma^{-1} \,.
\label{errors}
\end{equation}
The correlation coefficients between two parameters $\theta^a$ and
$\theta^b$ are given by
\begin{equation}
c_{ab} = \Sigma^{ab}/\sqrt{\Sigma^{aa}\Sigma^{bb}} \,.
\label{correlations}
\end{equation}
The assumption of large SNR is well justified for MBH binaries,
which can be observed with high SNR out to very large distances. 

To determine how accurately {\em LISA} can measure source locations
and luminosity distances we perform Monte Carlo simulations using a
population of sources across the sky. We consider in detail two
classes of systems. The first are NS-IMBH binaries with a NS mass
$M_{\rm NS}=1.4~M_\odot$ and a BH mass in the range
$[400-10^4]~M_\odot$, a typical target system used to place bounds on
the BD parameter. According to current estimates, detection rates for
$10-10^2~M_\odot$ BHs spiralling into IMBHs of mass
$10^2-10^3~M_\odot$ are expected to be very low \cite{cliffrates}. 
In order to apply the Fisher matrix formalism in a consistent way, we
assume that these NS-IMBH binaries,
rare though they may be, are detected with a single-detector
SNR $\rho_I=10$. The second are MBH binaries. In this case we
first consider equal-mass binaries with total {\it measured} mass in
the range $[2\times 10^4-2\times 10^7]~M_\odot$ at distance
$D_L=3$~Gpc -- typical systems to put bounds on massive graviton
theories (Sec.~\ref{results}). Then we study parameter estimation for
MBH systems located at different cosmological redshifts and having
mass $(10^6+10^6)~M_\odot$ and $(10^7+10^7)~M_\odot$ {\it as measured
in the source rest frame} (Sec.~\ref{cosmology}).

For each of these systems we distribute $10^4$ sources over sky
position and orientation, described, in the solar system barycentric
frame, by angles $(\bt_S,\bph_S)$ and $(\bt_L,\bph_L)$, respectively.
We randomly generate the angles $\bph_S$,
$\bph_L$ in the range $[0,2 \pi]$ and $\mu_S = \cos \bt_S$, $\mu_L =
\cos \bt_L$ in the range $[-1,1]$ (see BBW for details). When the
waveform contains a large number of highly correlated parameters,
computing the inverse of the Fisher matrix can be numerically
difficult. The method we used to check the robustness of our results
is described in Appendix~B of BBW.

\section{Parameter estimation and bounds on alternative theories of gravity}
\label{results}

In BBW we carried out a set of Monte Carlo simulations for a MBH
binary of mass $(10^6+10^6)~M_\odot$ at fiducial distance
$D_L=3$~Gpc. We showed that our results are compatible with results
obtained using a simpler approach~\cite{willyunes}, in which 
we don't take into account \lisa's orbital motion and we 
average over pattern functions. 

\begin{figure*}
\begin{center}
\begin{tabular}{cc}
\epsfig{file=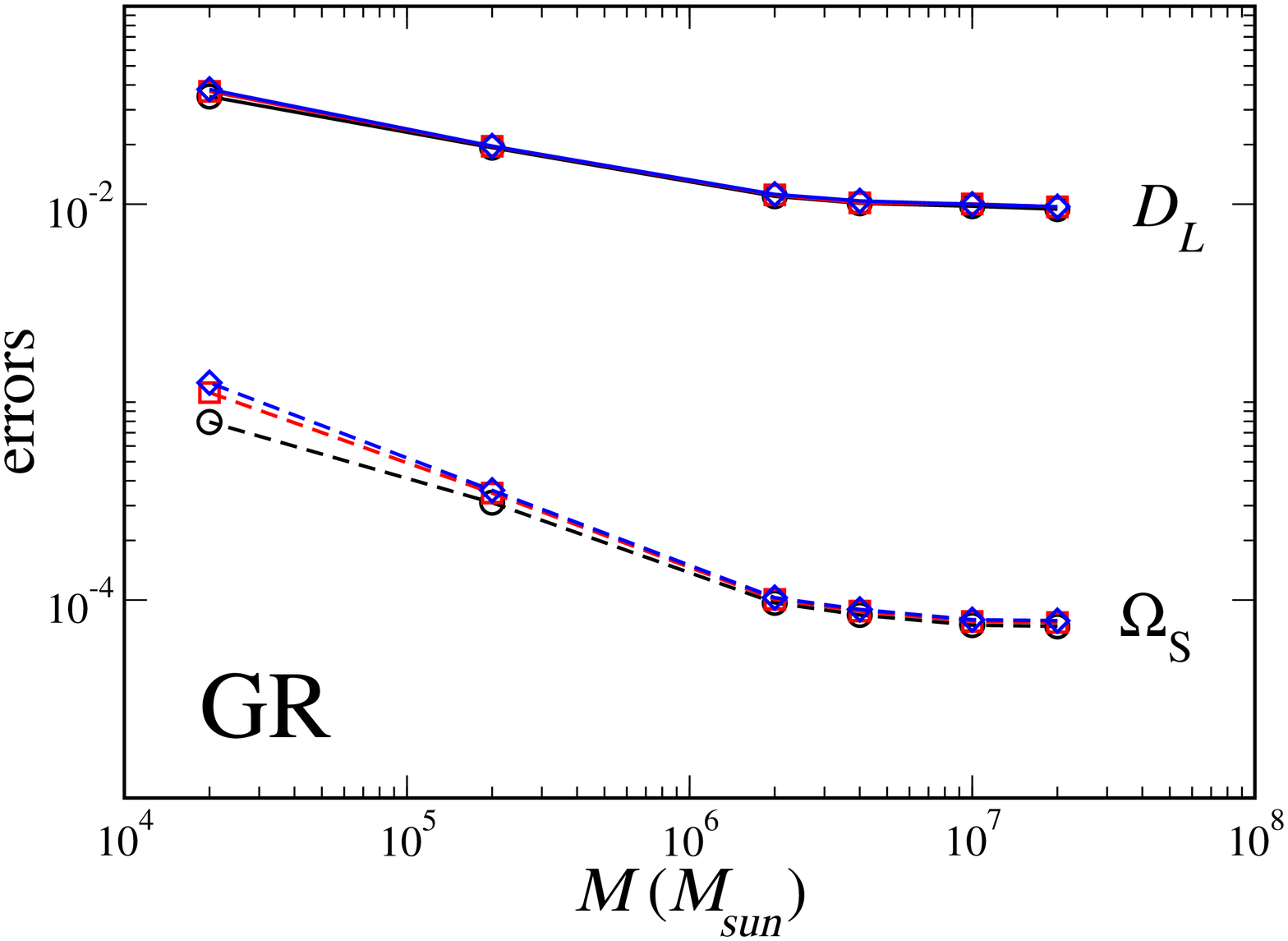,width=5cm,angle=-90}&
\epsfig{file=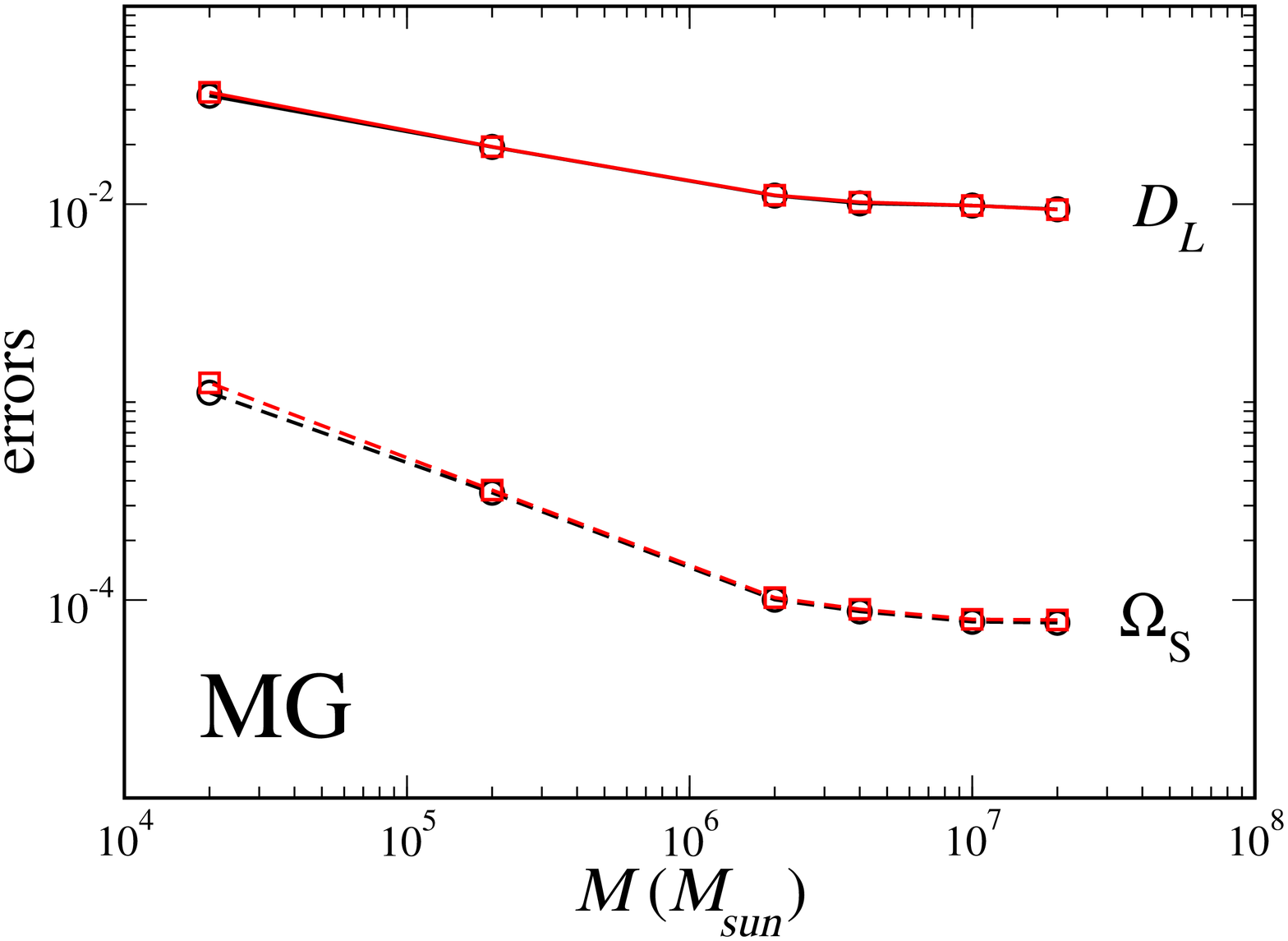,width=5cm,angle=-90}\\
\epsfig{file=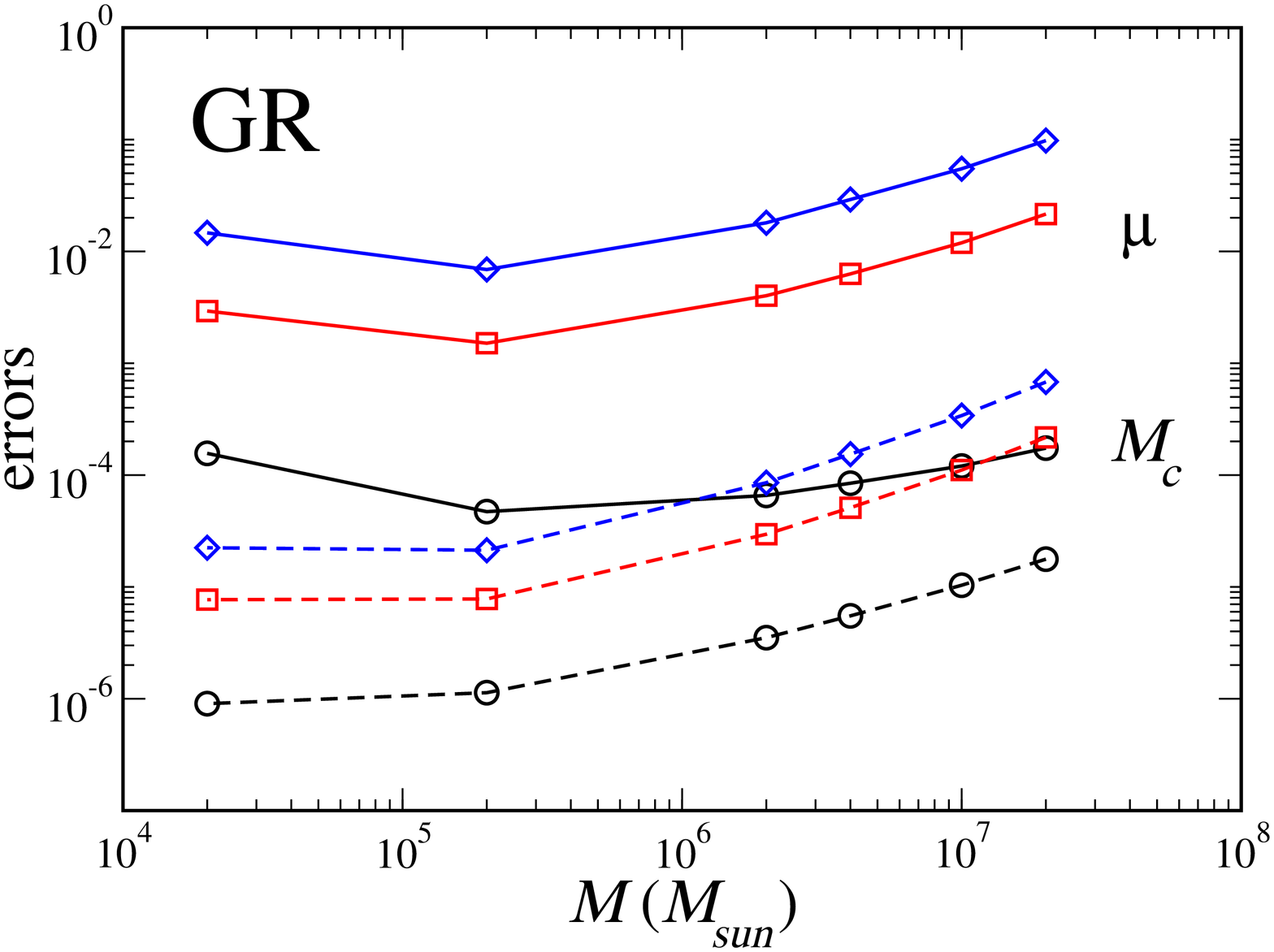,width=5cm,angle=-90}&
\epsfig{file=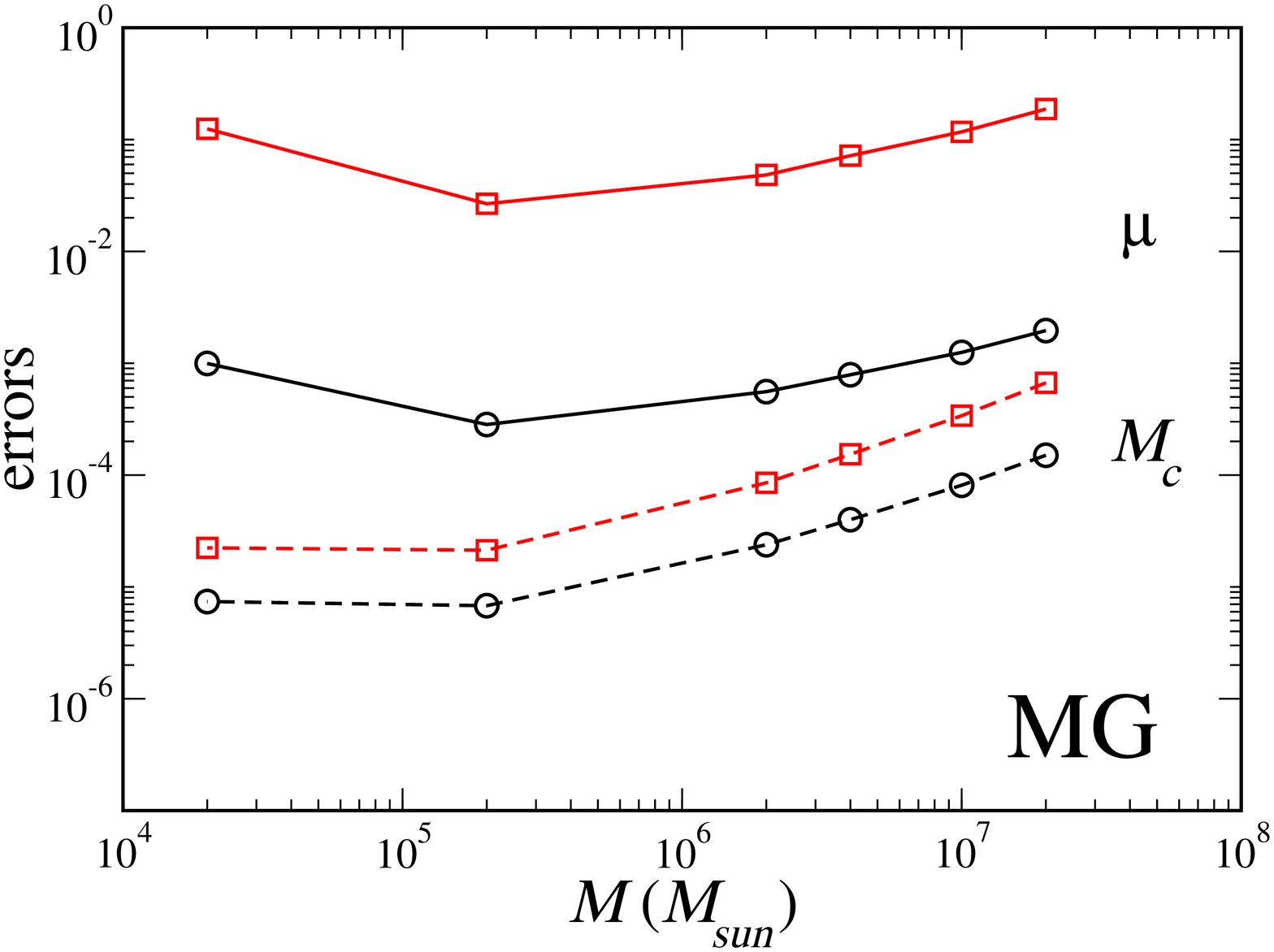,width=5cm,angle=-90}\\
\end{tabular}
\caption{Top: average errors on $\Omega_S$ (dashed lines) and $D_L$
(solid lines). Bottom: average errors on ${\cal M}$ (dashed lines) and
$\mu$ (solid lines). Left panel refers to GR, right panel to massive
graviton theories; errors are given as a function of the total mass
for equal mass MBH binaries. We assume the \lisa~noise curve can be
trusted down to $f_{\rm low}=10^{-5}$~Hz.  Black lines are computed
omitting spin terms, red lines include spin-orbit terms, blue lines
include both spin-orbit and spin-spin terms.
\label{BBWexpand}}
\end{center}
\end{figure*}

For this paper, we performed a more extensive set of Monte
Carlo simulations, spanning a wider range of binary masses. Results
from these simulations are shown in Fig.~\ref{BBWexpand}.  We consider
nonspinning equal-mass MBHs at fiducial distance $D_L=3$~Gpc, and we
investigate the dependence of mass, distance and angular resolution
determinations on the binary's total mass. From the plots on the first
row we see that distance and angular resolution determinations are
largely independent of the inclusion of spin terms (and even of the
massive-graviton terms) in the phase. Mass determinations are more
sensitive to the inclusion of spin terms and to the theory of gravity
we are considering. Distance errors and angular resolution generally
decrease with the binary's mass. On the contrary, mass errors have a
relative minimum in the given mass range. This is because the
measurement of mass parameters depends on an accurate modeling of the
phase of the waveform: when the MBH mass is large the binary's
frequency becomes low, the binary spends a significant fraction of
time out of the \lisa~sensitivity band, and the accuracy of mass
determinations decreases.

We also explored the dependence of parameter estimation on the
signal's ending frequency. Instead of using the Schwarzschild
Innermost Stable Circular Orbit (ISCO) as a conventional ending
frequency of the signal, we looked at the effect of truncating the
signal at the Minimum Energy Circular Orbit (MECO; see Sec.~IIB of
\cite{BCV} for a discussion), consistently computed at 2PN and
assuming dynamically small spins ($\beta=\sigma=0$). We found that
using the MECO instead of the ISCO has a completely negligible effect
on distance and angular position determinations. The effect on mass
determinations is completely negligible for total masses $M$ smaller
than $\sim 10^6 M_\odot$; even for values of $M$ larger than this,
using the MECO instead of the ISCO only results in slightly smaller
(by factors $\sim 2$ or so) errors. However, the effect of using the
MECO could be non-negligible if the spins are large.

\begin{figure*}
\begin{center}
\begin{tabular}{cc}
\epsfig{file=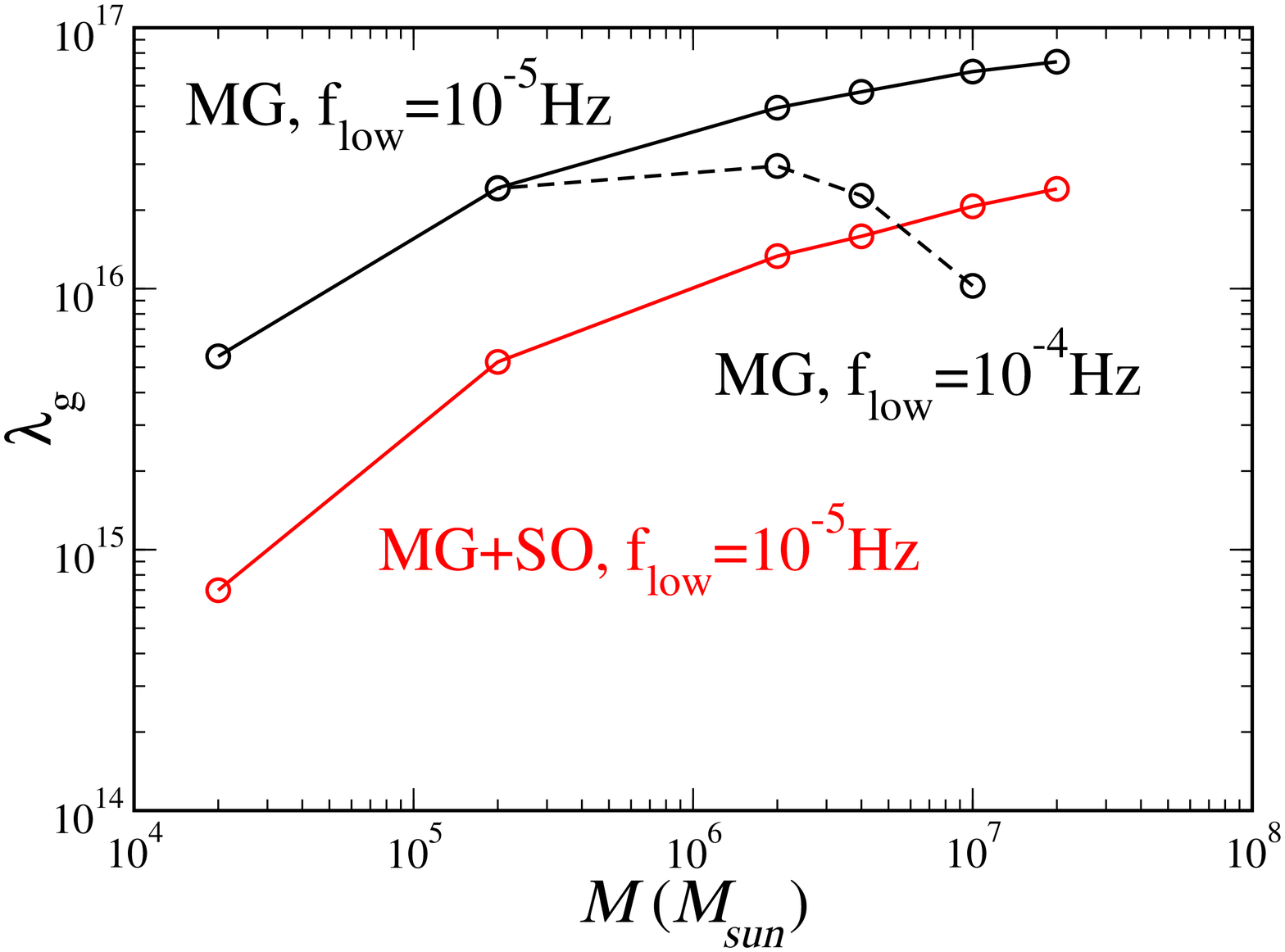,width=5cm,angle=-90}&
\epsfig{file=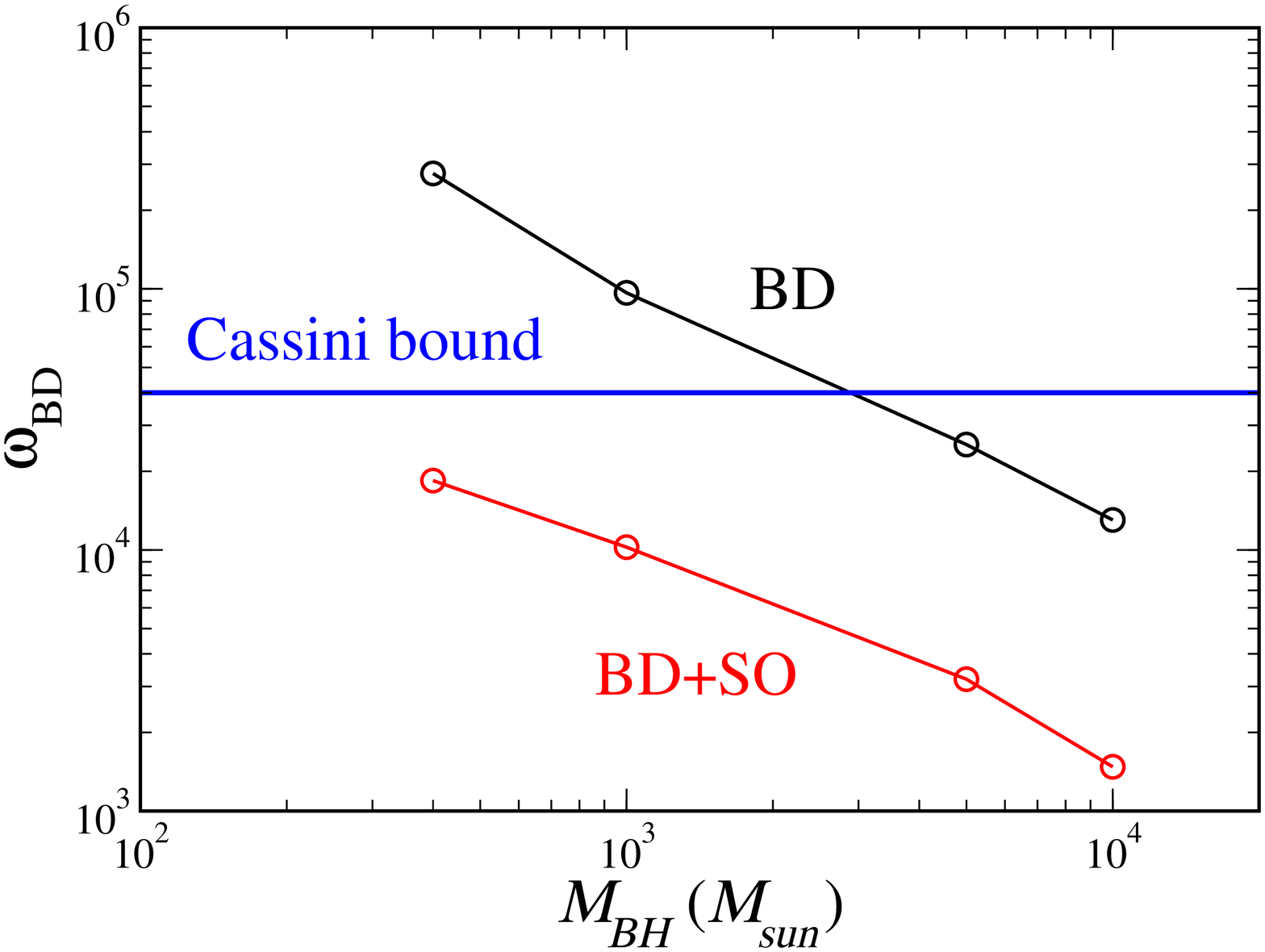,width=5cm,angle=-90}\\
\end{tabular}
\caption{Left: average bounds on the graviton wavelength as a function
of total mass for equal mass MBH binaries. Right: bound on the BD
parameter from a NS-IMBH binary as a function of the IMBH mass. The
horizontal blue line corresponds to the Cassini bound \cite{bertotti}.
Black lines are computed omitting spin terms; red lines include
spin-orbit terms as well. The dashed line shows that the bound on
$\lambda_g$ is reduced if the \lisa~noise curve can only be trusted
down to $f_{\rm low}=10^{-4}$~Hz.
\label{bounds}}
\end{center}
\end{figure*}

In Fig.~\ref{bounds} we look at bounds to be placed on i) the graviton
Compton wavelength $\lambda_g$, using a sample of equal-mass MBHs at
fiducial distance $D_L=3$~Gpc (left panel), and ii) the Brans-Dicke
parameter $\omega_{BD}$, using a sample of NS-IMBH binaries with
single-detector signal-to-noise ratio equal to 10, as a function of
the IMBH mass (right panel).  These results confirm qualitatively (and
improve quantitatively) results previously obtained by Will and
collaborators. Those authors adopted a somewhat cruder \lisa~model,
that does not take into account \lisa's orbital motion
\cite{willST,willgraviton,scharrewill,willyunes}. Fig.~\ref{bounds}
shows that understanding the low-frequency {\em LISA} noise is
important to set bounds on the graviton mass using high-mass MBH
binaries. We will see in Sec.~\ref{cosmology} that an understanding of
the low frequency noise is also important to use {\em LISA} in a
cosmological context.

\begin{figure*}
\begin{center}
\begin{tabular}{cc}
\epsfig{file=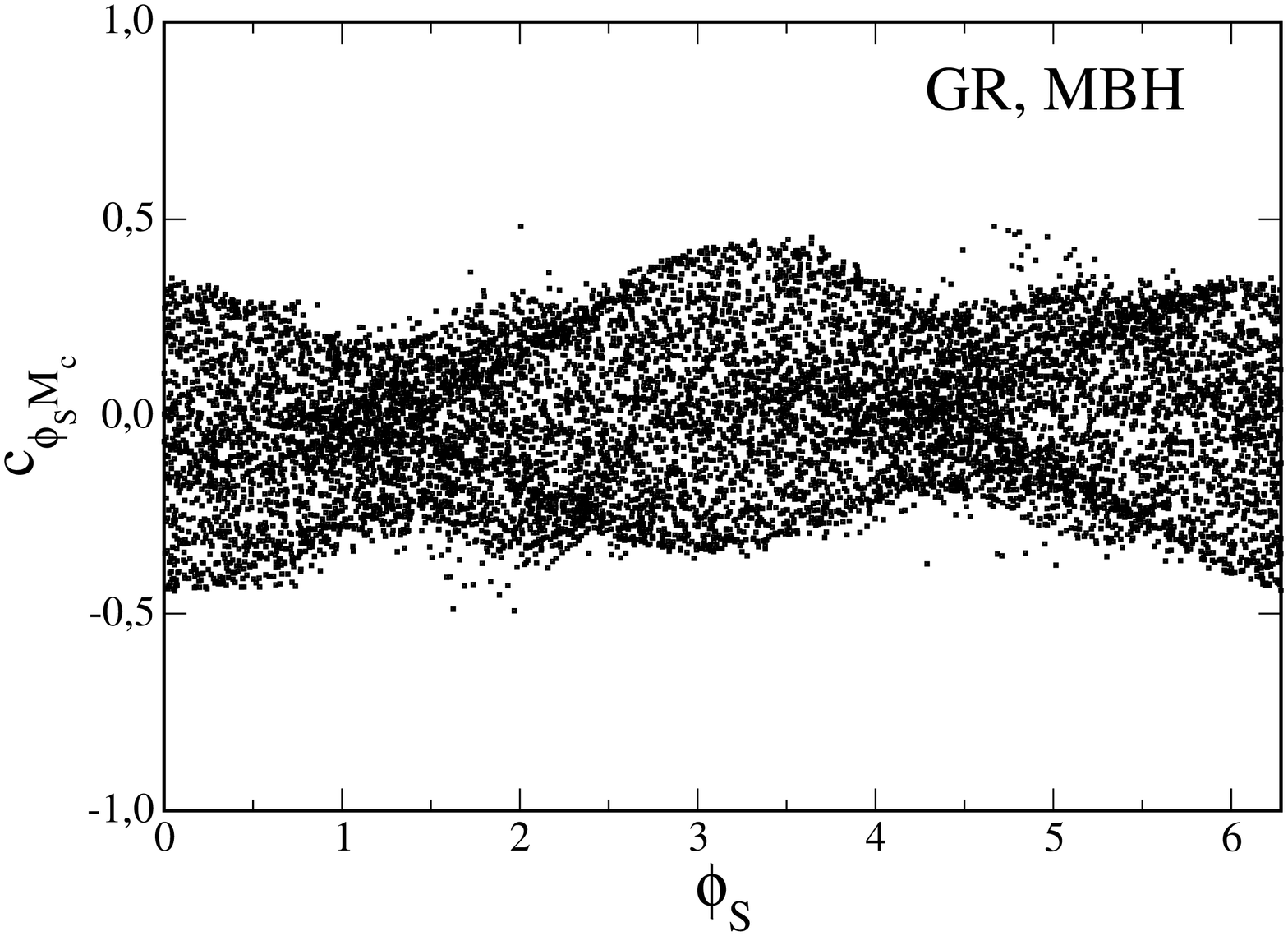,width=5cm,angle=-90}&
\epsfig{file=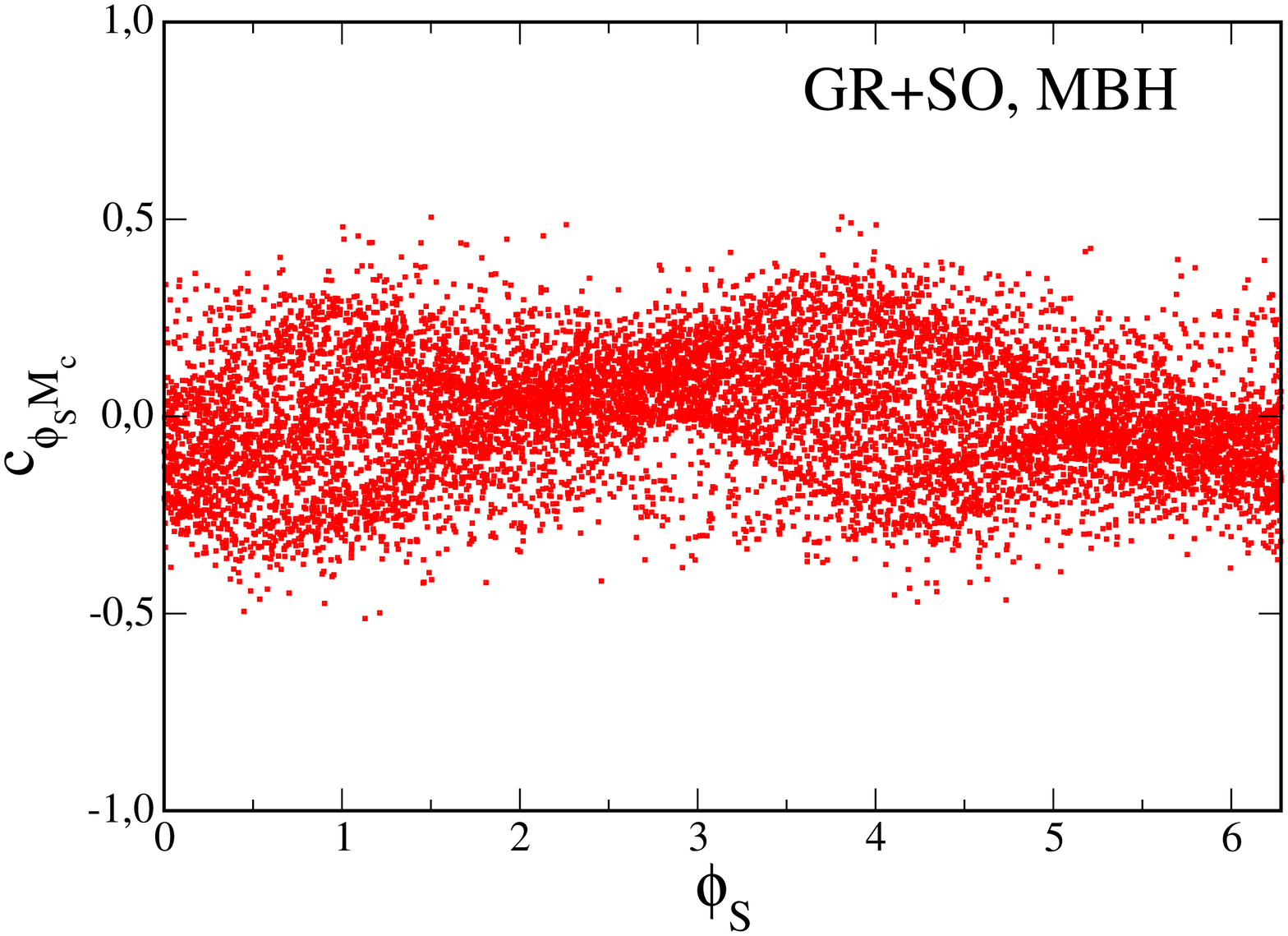,width=5cm,angle=-90}\\
\epsfig{file=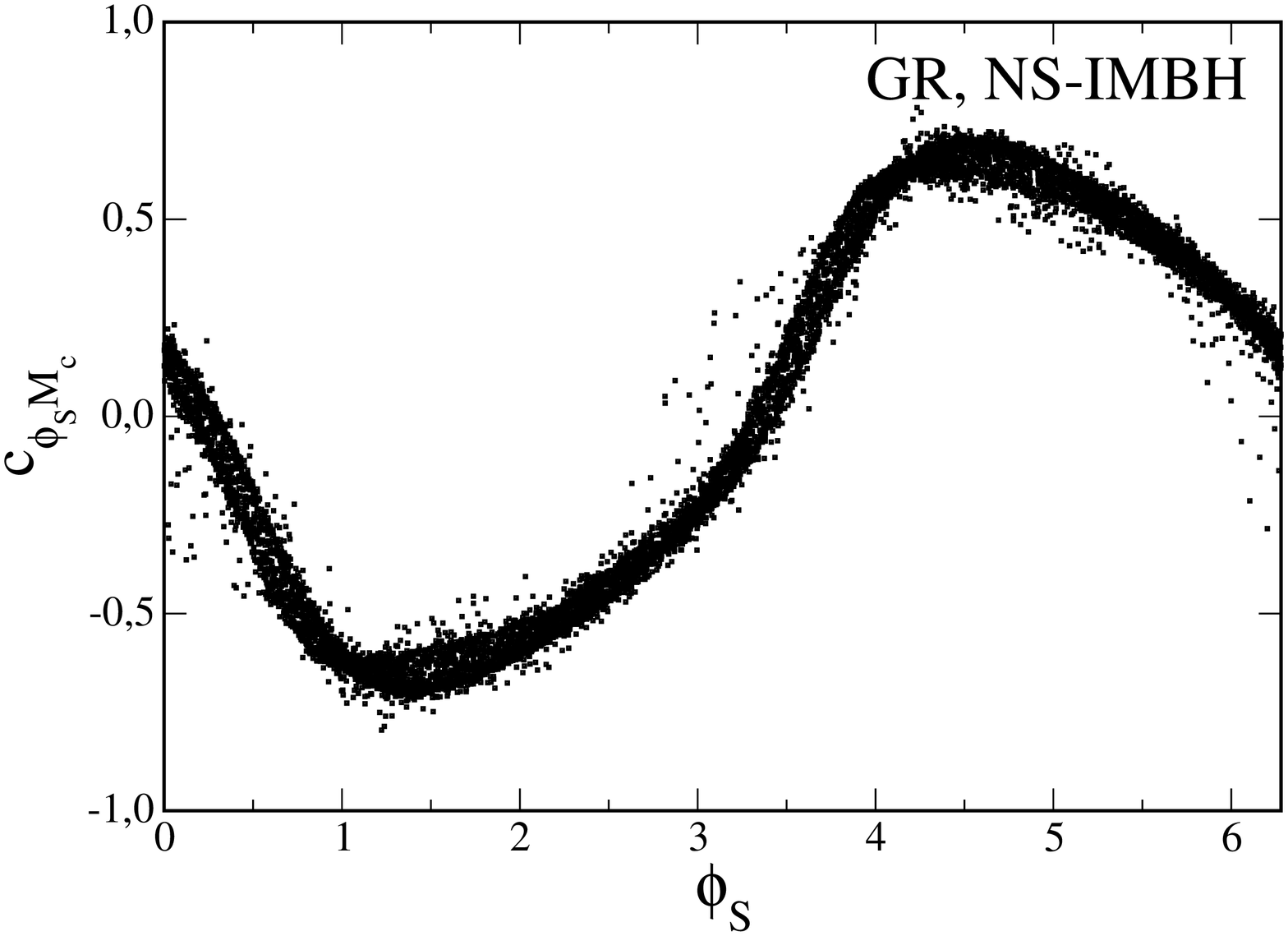,width=5cm,angle=-90}&
\epsfig{file=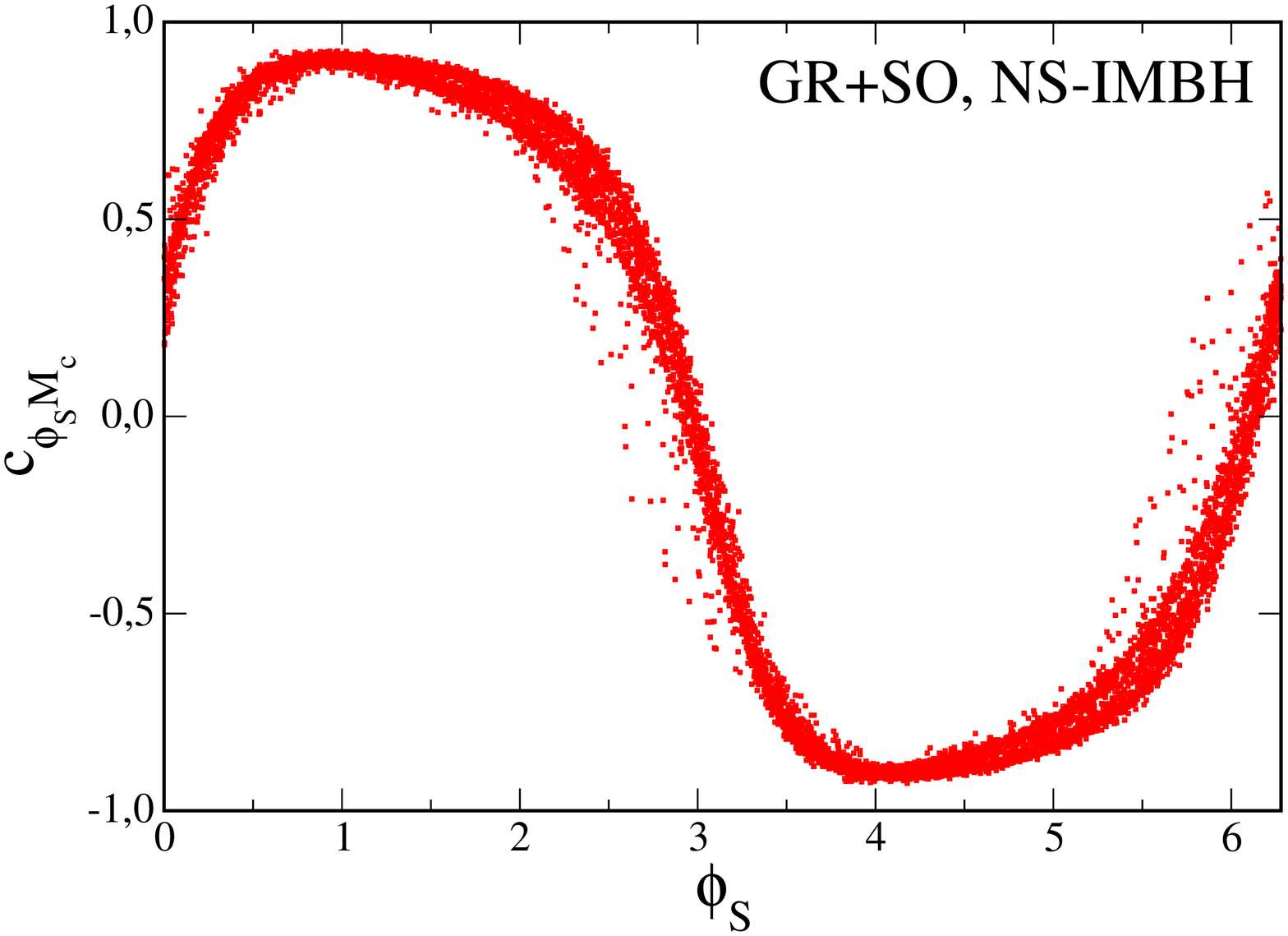,width=5cm,angle=-90}\\
\end{tabular}
\caption{Scatter plot of the correlations $c_{\bar \phi_S {\cal
M}}(\bar \phi_S)$ in GR. From left to right we observe how
correlations change when we add spins. The top row corresponds to a
$(10^6+10^6)~M_\odot$ MBH binary at $D_L=3$~Gpc; the bottom row refers
to a $(1.4+10^3)~M_\odot$ NS-IMBH system observed with single-detector
SNR equal to 10.
\label{corrs}}
\end{center}
\end{figure*}

In BBW we found that \lisa's angular resolution degrades when we
include spin terms in the NS-IMBH case, but it is essentially
unaffected in the MBH case.  To explore this further, in
Fig.~\ref{corrs} we show a scatter plot (out of $10^4$ binaries) of
the correlation $c_{\bar \phi_S {\cal M}}$ between the azimuthal angle
$\bar \phi_S$ (describing the source position in the sky with respect
to the Solar System barycenter) and the chirp mass ${\cal M}$
(dominating the phase evolution of the gravitational waveform), as a
function of $\bar \phi_S$, in the two cases. The correlation has a
``clean'' periodic behavior in the NS-IMBH case; in the MBH-MBH case
$c_{\bar \phi_S {\cal M}}$ is {\it not} a clean periodic function of
the binary's azimuthal angle $\bar \phi_S$, showing some beating
phenomena. An important feature is that, when we include spin terms,
the maximum correlation between mass and angular position clearly
grows in the NS-IMBH case, but not in the MBH-MBH case. This is
consistent with the fact that the angular resolution depends on
spin couplings in the NS-IMBH case, but it doesn't in the MBH-MBH
case. 

\section{Probing the merger history of supermassive black holes with LISA}
\label{cosmology}

Astrophysical estimates of MBH coalescence rates depend on many
complex physical phenomena. The study of MBH binary evolution in
galaxy mergers is a very active field of research \cite{MM}. A class
of models predicts that the {\em LISA} mission should be able to {\it
detect} at least $\sim 10$ MBH coalescence events out to large values
of the redshift, and that most of these events would originate at
$z\sim 2-6$~ \cite{sesana2,RW,menou}. Research in this field is very
active and the estimates are still quite uncertain: a partial list of
references includes \cite{H94,sesana1,EINS,ITS,MSE,KZ}. 

To use {\em LISA} as a cosmological probe it would be desirable to
have an accurate measurement of the binary parameters, and not just a
simple detection. In this context the redshift $z$ can be large, and
the distinction between observed masses and masses as measured in the
source rest frame -- given by the simple rescaling ${M}_{\rm observed}
= (1+z) {M}_{\rm source}$, where $z$ is the cosmological redshift --
is important\footnote{Up to this point we denoted by ${\cal M}$ and
$M$ the {\it observed} chirp and total masses.}. {\em LISA} can only
measure redshifted combinations of the intrinsic source parameters
(masses and spins), so it cannot measure the redshift $z$ to the
source. But if the MBH masses, luminosity distances and angular
resolution are determined with sufficient accuracy, {\em LISA} can
still provide information on the growth of structures at high
redshift.  If cosmological parameters are known, the relation between
luminosity distance and redshift $D_L(z,H_0,\Omega_M,\Omega_\Lambda)$
(where $H_0=72$~km~s$^{-1}$~Mpc$^{-1}$ is Hubble's constant and we
assume a zero--spatial-curvature Universe with $\Omega_\kappa=0,~
\Omega_\Lambda + \Omega_M =1$, according to the present observational
estimates) can be inverted to yield
$z(D_L,H_0,\Omega_M,\Omega_\Lambda)$. Then {\em LISA} measurements of
the luminosity distance can be used to obtain intrinsic BH masses as a function
of redshift, thus constraining hierarchical merger
scenarios~\cite{SH}. Alternatively: if we can obtain the binary's
redshift by some other means, e.g. from an electromagnetic
counterpart, then {\em LISA} measurements of $D_L$ can be used to
improve our knowledge of the cosmological
parameters~\cite{schutz,markovic,HH}.

These exciting applications depend, of course, on {\em LISA}'s
measurement accuracy at large redshifts. Here, as in BBW, we look at
the redshift dependence of measurement errors for two representative
MBH binaries having masses $(10^6+10^6) M_\odot$ and $(10^7+10^7)
M_\odot$ {\em as measured in the source rest frame}.  We also assume
that the {\em LISA} noise can be extrapolated down to $f_{\rm
low}=10^{-5}$~Hz; more conservative assumptions on $f_{\rm low}$ could
significantly affect our conclusions (see~\cite{HH,Baker,BBW}). We
compute the errors performing Monte Carlo simulations of $10^4$
binaries for different values of the redshift and then averaging over
all binaries. 

\begin{figure*}
\begin{center}
\begin{tabular}{cc}
\epsfig{file=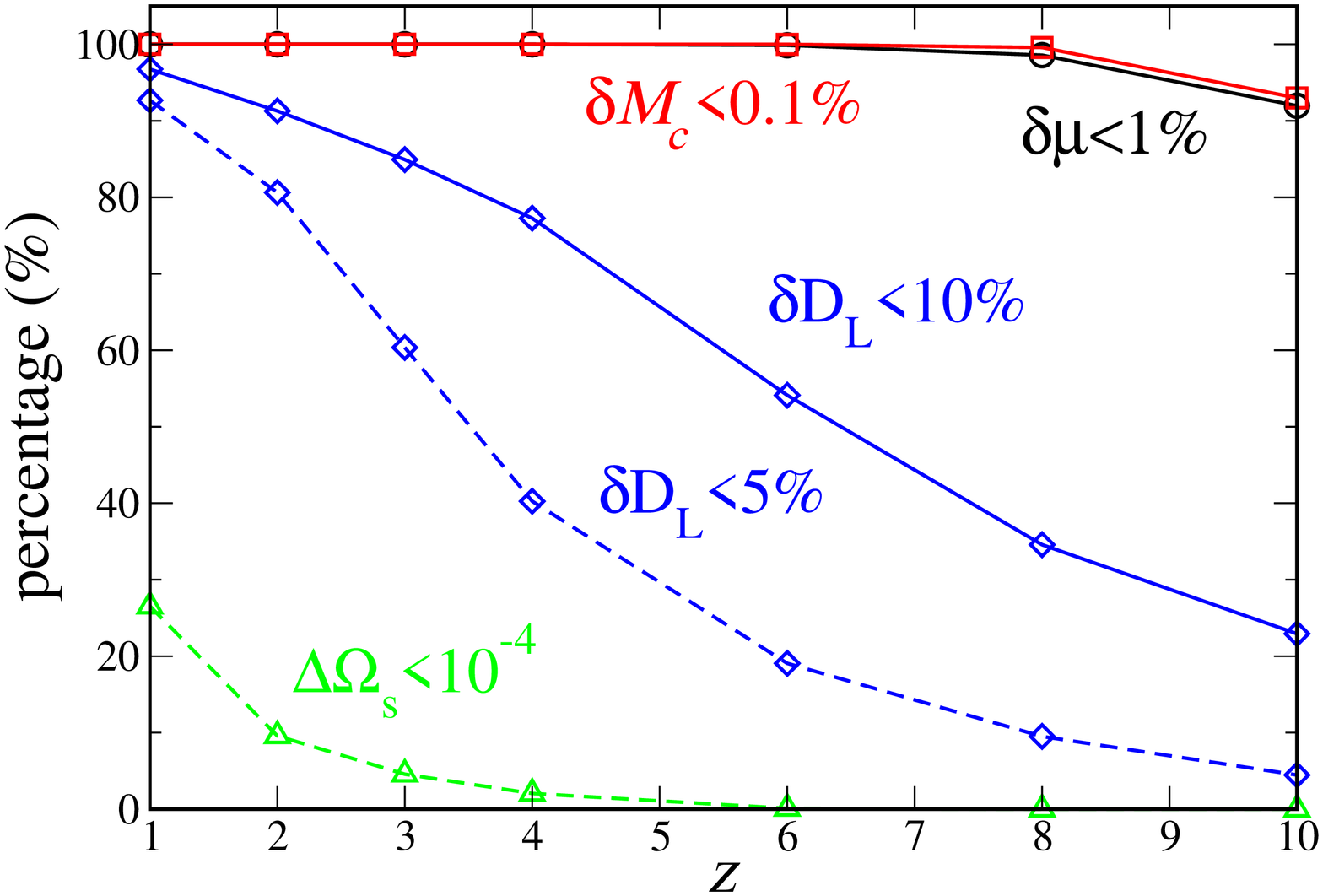,width=5cm,angle=-90}&
\epsfig{file=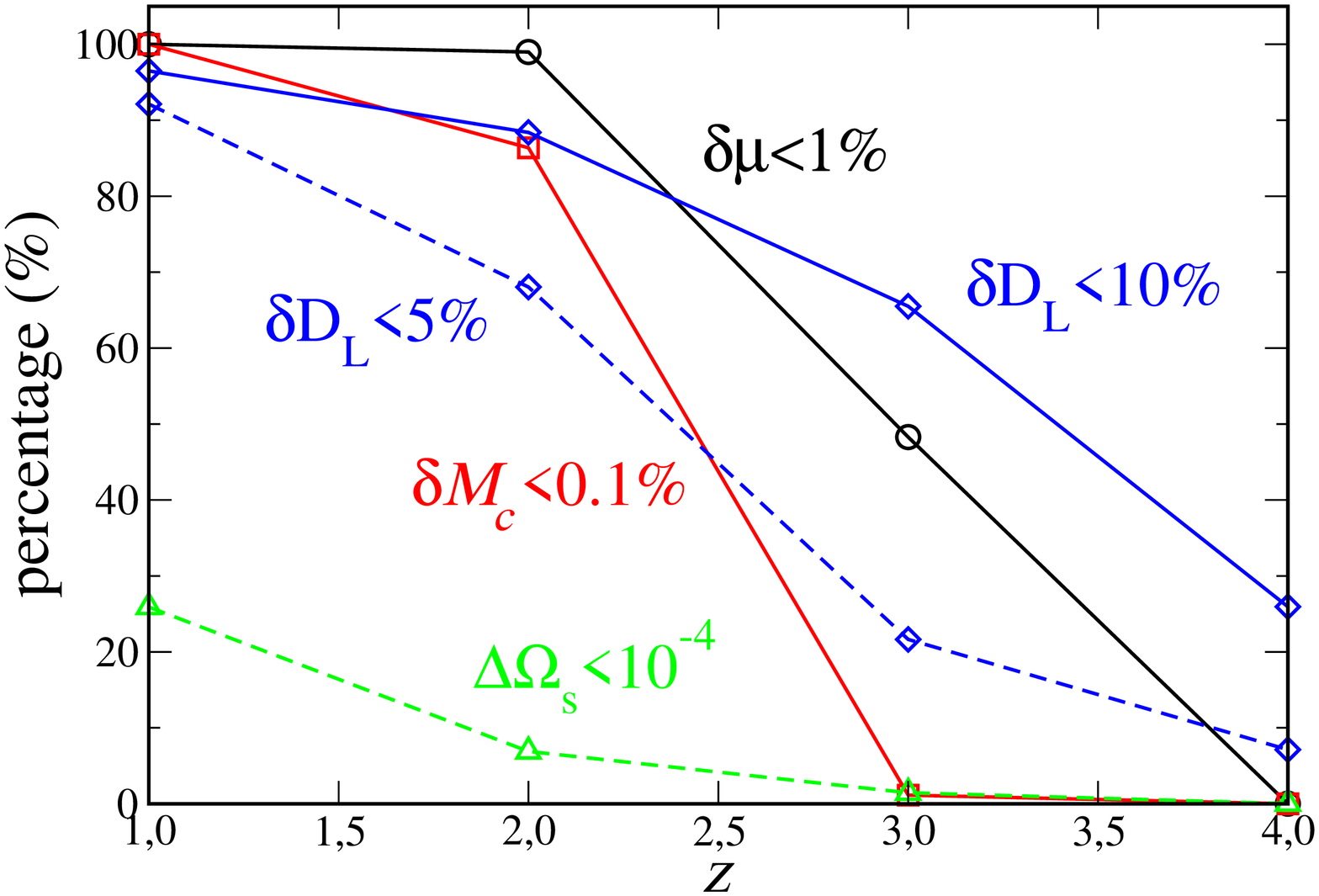,width=5cm,angle=-90}
\end{tabular}
\caption{Percentage of binaries for which: 
$\delta \mu = \Delta \mu/\mu<1 \%$ (black circles),
$\delta {\cal M}=\Delta {\cal M}/{\cal M}<0.1 \%$ (red squares;
$\delta {\cal M }=\Delta {\cal M}/{\cal M}<1 \%$ in all cases considered),
$\delta D_L = \Delta D_L/D_L<10 \%$ (blue diamonds),
$\delta D_L = \Delta D_L/D_L<5 \%$ (blue diamonds, dashed lines),
$\Delta \Omega_S<10^{-4}$ (green triangles).
On the left we consider a BH-BH binary of mass $(10^6+10^6)~M_\odot$,
on the right a BH-BH binary of mass $(10^7+10^7)~M_\odot$. 
We assume $f_{\rm low}=10^{-5}$~Hz.
\label{perc}}
\end{center}
\end{figure*}

As discussed in Sec.~\ref{results}, distance determination and angular
resolution for MBHs are essentially independent of the inclusion of
spin terms. The (average) relative error on $D_L$ for the $(10^6+10^6)
M_\odot$ system is $\sim 2 \%$ at $z=1$, $\sim 5 \%$ at $z=2$ and
$\sim 11 \%$ at $z=4$.  This reduction in accuracy is due to the fact
that the signal spends less and less time in band as the redshift is
increased. We only consider values of the redshift such that the
binary spends at least one month in band before coalescing: for the
case $(10^7+10^7) M_\odot$, this corresponds to $z\sim 4$. The
distance determination error for this high-mass binary grows quite
rapidly, being $\sim 2 \%$ at $z=1$, $\sim 6 \%$ at $z=2$ and $\sim 21
\%$ at $z=4$. {\em LISA}'s angular resolution is rather poor even at
small redshifts, and it rapidly degrades for sources located farther
away, the degradation being more pronounced for higher-mass
binaries. Better distance determinations can be obtained if we are
lucky enough to locate the source in the sky by some other means: for
example, associating the gravitational wave event with an
electromagnetic counterpart. In this case angles and distance would be
decorrelated, allowing order of magnitude improvements in the
determination of $D_L$~\cite{HH}. For $(10^6+10^6) M_\odot$ general
relativistic nonspinning binaries, a least-square fit of mass and
distance errors in the interval $z\in [1,10]$ yields:
\bea
\Delta {\cal M}/{\cal M}&=&
(-0.1148+0.7236 z+0.5738 z^2)\times 10^{-5}\,, \nn\\
\Delta \mu/\mu&=&
(-0.61431+1.9018 z+0.43721 z^2)\times 10^{-4}\,, \nn\\
\Delta D_L/D_L&=&
(-0.65651+2.6935 z+0.061595 z^2)\times 10^{-2}\,. \nn
\eea
It is important to stress that in our discussion above we refer to
{\it average} errors from a population of $10^4$ sources randomly
distributed across the sky. In a single ``lucky'' detection, parameter
estimation errors could be much smaller than the average. The most
useful quantity to measure \lisa's cosmological capabilities is
perhaps the {\it percentage of binaries whose parameters can be
measured within some given accuracy}, as a function of $z$. This
percentage is shown in Fig.~\ref{perc}.

Upper limits on measurement errors of the various parameters have been
chosen in order to answer the following questions. Out to which value
of $z$ can we determine the chirp mass to better than $0.1~\%$, and
the reduced mass to better than $1~\%$?  Out to which $z$ does {\em
LISA} beat the present uncertainty ($\sim 10 \%$) on cosmological
parameters, allowing us to determine the luminosity distance to the
source? Which fraction of binaries can be observed with an angular
resolution error $\Delta \Omega_S<10^{-4}$, roughly corresponding to
the angular diameter of the Moon (and the Sun) as seen from the Earth?

The answer to these questions depends of course on the binary's mass
and on the low-frequency sensitivity of {\em LISA}. Following BBW, in
our calculations we assume that the noise curve can be extrapolated
down to $f_{\rm low}=10^{-5}$~Hz and we consider two representative
binaries with mass $(10^6+10^6)~M_\odot$ and $(10^7+10^7)~M_\odot$,
respectively.

The main conclusions to be drawn from Fig.~\ref{perc} are the
following.  (i) The fractional error on the chirp mass can be better
than $0.1~\%$ out to $z=10$ and beyond for a $(10^6+10^6)~M_\odot$
binary, but only out to $z\simeq 2$ for a $(10^7+10^7)~M_\odot$
binary.  (ii) In the absence of spins, the fractional error on the
reduced mass can be better than $\sim 1~\%$ out to $z=10$ and beyond
for a $(10^6+10^6)~M_\odot$ binary, but only out to $z\simeq 2$ for a
$(10^7+10^7)~M_\odot$ binary. (iii) For $\sim 80~\%$ of observed
binaries, {\em LISA}'s distance determination will be better than
$\sim 10 \%$ (beating present errors on cosmological parameters) out
to $z\simeq 4$ for a $(10^6+10^6)~M_\odot$ binary, and out to $z\simeq
2$ for a $(10^7+10^7)~M_\odot$ binary.  (iv) \lisa's angular
resolution is comparatively quite poor: only a small fraction of
binaries can be observed with $\Delta \Omega_S<10^{-4}$ at any
redshift.

Unimportant as they are for distance determination and angular
resolution, spin effects have a dramatic impact on mass measurement
accuracy. In BBW we showed that when we include spin-orbit and
spin-spin terms, the error on the chirp mass is still pretty small: it
only becomes larger than a few percent for massive ($M\sim 2\times
10^7 M_\odot$) binaries located at $z>2$. However, non-precessional
spin effects seriously limit our ability to measure the reduced mass.
When we include both spin-orbit and spin-spin terms, the reduced mass
error becomes larger than 5 \% at all redshifts even for ``low mass'',
$(10^6+10^6)~M_\odot$ binaries.  Our conclusions on mass
determinations for spinning binaries could change when we consistently
include {\it precessional} spin effects. These effects can induce
modulations in the waveform, possibly improving the mass measurements
in a significant way~\cite{vecchio}. Therefore the study of precession
is very important to assess {\em LISA}'s ability to measure MBH masses
in galactic mergers; we plan to return to this study in the future.

\section*{Acknowledgments}
This work was supported in part by the National Science Foundation
under grant PHY 03-53180.

\vskip .5cm




\begin{thebibliography}{999}
\bibitem{danzmann} Danzmann K for the {\em LISA} Science Team
1997
\CQG {\bf 14} 1399 
%
\bibitem{BBW}
Berti E, Buonanno A and Will C M 
2005
\PR D {\bf 71} 084025
({\em Preprint} gr-qc/0411129)
%
\bibitem{willST} 
Will C M 
1994
\PR D {\bf 50} 6058 
({\em Preprint} gr-qc/9406022)
%
\bibitem{willgraviton} 
Will C M
1998
\PR D {\bf 57} 2061 
({\em Preprint} gr-qc/9709011)
%
\bibitem{scharrewill} 
Scharre P D and  Will C M  
2002
\PR D {\bf 65}  042002 
({\em Preprint} gr-qc/0109044)
%
\bibitem{willyunes} 
Will C M  and Yunes N   
2004 
\CQG {\bf 21} 4367 
({\em Preprint} gr-qc/0403100)
%
\bibitem{damourfarese} 
Damour T and Esposito-Far\`ese G 
1998
\PR D {\bf 58} 042001 
({\em Preprint} gr-qc/9803031)
%
\bibitem{CC} 
Cutler C 
1998
\PR D {\bf 57} 7089 
({\em Preprint} gr-qc/9703068)
%
\bibitem{SH} 
Hughes S A  
2002
{\em Mon. Not. Roy. Astron. Soc.} {\bf 331} 805 
({\em Preprint} astro-ph/0108483)
%
\bibitem{seto} 
Seto N 
2002
\PR D {\bf 66} 122001 
({\em Preprint} gr-qc/0210028)
%
\bibitem{vecchio} 
Vecchio A 
2004
\PR D {\bf 70} 042001 
({\em Preprint} astro-ph/0304051)
%
\bibitem{HM} 
Hughes S A  and Menou K 
2005
{\em Astrophys. J.} {\bf 623} 689
({\em Preprint} astro-ph/0410148)
%
\bibitem{BH} 
Bender P L  and Hils D  
1997
\CQG {\bf 14} 1439 
%
\bibitem{BC} 
Barack L and Cutler C 
2004 
\PR D {\bf 69} 082005
({\em Preprint} gr-qc/0310125)
(See also L Barack and C Cutler, gr-qc/0310125 v3, where an
erroneous factor of 3/4 in the instrumental noise is corrected)
%
\bibitem{SCG} The Sensitivity Curve Generator was originally written
by Shane Larson and may be found online at
http://www.srl.caltech.edu/~shane/sensitivity/MakeCurve.html
%
\bibitem{poissonwill} 
Poisson E and Will C M 
1995
\PR D {\bf 52} 848 
({\em Preprint} gr-qc/9502040)
%
\bibitem{KKS}
Krolak A, Kokkotas K D and Sch\"afer G 
1995
\PR D {\bf 52} 2089 
({\em Preprint} gr-qc/9503013)
%
\bibitem{bertotti}
Bertotti B, Iess L and Tortora P 
2003
{\em Nature} {\bf 425} 374 
%
\bibitem{HH} 
Hughes S A  and Holz D E  
2003
\CQG {\bf 20} S65 
({\em Preprint} astro-ph/0212218)
%
\bibitem{Baker} 
Baker J and Centrella J 
2004
({\em Preprint} astro-ph/0411616)
%
\bibitem{sesana2}
Sesana A, Haardt F, Madau P and Volonteri M 
2004
{\em Preprint} astro-ph/0409255 ({\em Astrophys. J.} at press)
%
\bibitem{RW}
Rhook K J  and Wyithe J S B 
2005
{\em Preprint} astro-ph/0503210 ({\em Mon. Not. Roy. Astron. Soc.} at
press)
%
\bibitem{menou}
Menou K, Haiman Z and Narayanan V K 
2001
{\em Astrophys. J.} {\bf 558} 535 
({\em Preprint} astro-ph/0101196)
%
\bibitem{prince}
Prince T A, Tinto M, Larson S L and Armstrong J W
2002
\PR D {\bf 66} 122002
({\em Preprint}  gr-qc/0209039)
%
\bibitem{blanchet1}
Blanchet L, Faye G, Iyer B R and Joguet B 
2002
\PR D {\bf 65} 061501 
({\em Preprint} gr-qc/0105099)
%
\bibitem{blanchet2} 
Blanchet L, Damour T, Esposito-Far\`ese G and Iyer B R 
2004
\PRL {\bf 93} 091101 
({\em Preprint} gr-qc/0406012)
%
\bibitem{willzaglauer}
Will C M and Zaglauer H W 
1989
{\em Astrophys. J.} {\bf 246} 366 
%
\bibitem{finn}
Finn L S  
1992
\PR D {\bf 46} 5236 
({\em Preprint} gr-qc/9209010)
%
\bibitem{FinnChernoff}
Finn L S and Chernoff D F  
1993
\PR D {\bf 47} 2198 
({\em Preprint} gr-qc/9301003)
%
\bibitem{CutlerFlanagan}
Cutler C and Flanagan \'E E  
1994
\PR D {\bf 49} 2658 
({\em Preprint} gr-qc/9402014)
%
\bibitem{cliffrates}
Will C M 
2004
{\em Astrophys. J.} {\bf 611} 1080 
({\em Preprint} astro-ph/0403644)
%
\bibitem{BCV}
Buonanno A, Chen Y and Vallisneri M 
2003
\PR D {\bf 67} 104025 
({\em Preprint} gr-qc/0211087)
%
\bibitem{MM}
Merritt D and Milosavljevi\'c M 
2004
{\em Preprint} astro-ph/0410364
\bibitem{H94}
Haehnelt M G  
1994
{\em Mon. Not. Roy. Astron. Soc.} {\bf 269} 199 
({\em Preprint} astro-ph/9405032)
%
\bibitem{sesana1}
Sesana A, Haardt F, Madau P and Volonteri M 
Astrophys. J. {\bf 611} 623 
2004
({\em Preprint} astro-ph/0401543)
%
\bibitem{EINS}
Enoki M, Inoue K T, Nagashima M and Sugiyama N 
2004 
{\em Astrophys. J.} {\bf 615} 19 
({\em Preprint} astro-ph/0404389)
%
\bibitem{ITS}
Islam R R, Taylor J E and Silk J 
2004
{\em Mon. Not. Roy. Astron. Soc.} {\bf 354} 629
({\em Preprint} astro-ph/0309559)
%
\bibitem{MSE}
Matsubayashi T, Shinkai H and Ebisuzaki T 
2004 
{\em Astrophys. J.} {\bf 614} 864 
%
\bibitem{KZ}
Koushiappas S M and Zentner A R 
2005
{\em Preprint} astro-ph/0503511
%
\bibitem{schutz}
Schutz B 
1986
{\em Nature} {\bf 323} 310 
%
\bibitem{markovic}
Markovic D 
1993
\PR D {\bf 48} 4738 

\end{thebibliography}
\end{document}